\documentclass[eclepsf,showpacs,preprintnumbers,amsmath,amssymb]{revtex4}

\usepackage{graphicx}
\usepackage{amsmath}
\usepackage{bm}
\usepackage{multirow}
\usepackage{comment}

\def\title#1{\begin{center}{\Large\bf #1}\end{center}}
\def\author#1{\vskip 5mm \begin{center}{#1}\end{center}}
\def\address#1{\begin{center}{\it #1}\end{center}}
\newcommand{\simgt}{\lower.5ex\hbox{$\; \buildrel > \over \sim \;$}}
\newcommand{\simlt}{\lower.5ex\hbox{$\; \buildrel < \over \sim \;$}}

\begin{document}

\title{First-order quantum correction to the Larmor radiation from a
moving charge in spatially homogeneous time-dependent electric field}

\author{Kazuhiro Yamamoto and Gen Nakamura}
\email[Email:]{kazuhiro@hiroshima-u.ac.jp}
 
\address{
Department of Physical Science, Hiroshima University,
Higashi-Hiroshima 739-8526,~Japan}

\begin{abstract}
First-order quantum correction to the 
Larmor radiation is investigated on the basis of the 
scalar QED on a homogeneous background of
a time-dependent electric field, which is a generalization 
of a recent 
work by Higuchi and Walker so as to be extended for 
an accelerated charged particle in a relativistic motion.
We obtain a simple approximate formula for the quantum 
correction in the limit of the relativistic motion
when the direction of the particle motion 
is parallel to that of the electric field.
\end{abstract}

\maketitle


\def\bfp{{\bf p}}
\def\bfk{{\bf k}}
\def\bfx{{\bf x}}
\def\bfA{{\bf A}}
\def\eta{{t}}
\def\bfv{{\bf v}}
\section{Introduction}
The Larmor radiation is the classical radiation from a charged
particle in an accelerated motion \cite{Jackson}. 
In a recent paper by Higuchi and Walker \cite{HW}, 
the quantum correction to the Larmor radiation is investigated on 
the basis of the scalar quantum electrodynamics (QED). 
In their approach, the mode function for the complex scalar field 
is constructed with the Wentzel-Kramers-Brillouin (WKB) approximation, 
in a form expanded with respect to $\hbar$. 
In a series of Higuchi and Martin's work \cite{HM,HMII,HMIII} 
(see also references therein), 
it has been well understood that the mode function reproduces the classical 
Larmor formula when the radiation energy is evaluated at the order of 
$\hbar^0$. The first-order 
quantum correction to the classical Larmor radiation is 
evaluated at the order of $\hbar$ in Ref.~\cite{HW}, though the 
investigation is limited to the non-relativistic motion of the 
charged particle.

In the present paper, we consider a simple generalization of Higuchi and 
Walker's work \cite{HW}, in order to investigate the case a relativistic 
motion of an accelerated charge.
Assuming a homogeneous but time-varying background of electric field, 
we derive a formula for the radiation energy of the order of $\hbar$, 
the first-order correction due to the quantum effect. 
This generalized formula is applicable to the accelerated charge in 
a relativistic motion, and we focus our investigation on the 
first-order quantum correction to the Larmor radiation in the limit 
of the relativistic motion.
This paper is organized as follows: In section 2, we present the 
general formula for the first-order quantum correction to the Larmor radiation.  
In section 3, we show that the formula reproduces
the same result obtained in Ref.~\cite{HW}, in the 
limit of the non-relativistic motion of the accelerated charge.
Then, an approximate formula in the limit of the relativistic motion is 
presented. Section 4 is devoted to summary and conclusions. 
In the appendix A, a brief derivation of the approximate formulas
is summarized. In the appendix B, we consider the validity of the
WKB approximation. 
Throughout this paper, we use units in which the velocity of light 
equals $1$, unless stated otherwise.


\section {formulation}
We consider the scalar QED with the action,
\begin{eqnarray}
S=\int d\eta d^3\bfx \left[
\left(D_\mu \phi\right)^\dagger D^\mu \phi-{m^2\over \hbar^2}\phi^\dagger\phi
-{1\over 4\mu_0}F_{\mu\nu}F^{\mu\nu}
\right],
\end{eqnarray}
where $D_\mu=\left({\partial/\partial x^\mu}+{ieA_\mu/\hbar}\right)$, 
$e$ and $m$ are the charge and the mass of the massive scalar field, 
respectively, and $\mu_0$ is the magnetic permeability of vacuum.
We work in the Minkowski spacetime, but consider the homogeneous electric 
background field ${\bf E}(\eta)$, which is related to the vector
potential by $\overline{A}_\mu=(0,{\bf A}(\eta))$ and 
$\dot{\bf A}(\eta)=-{\bf E}(\eta)$, where the dot denotes the
differentiation with respect to the time.
The equation of motion of the free scalar field yields
\begin{eqnarray}
&&\left({\partial^2 \over \partial \eta^2}+{(\bfp -e{\bfA}(t))^2+m^2\over \hbar ^2}
\right)\varphi_\bfp(\eta)=0,
\label{eqmod}
\end{eqnarray}
where $\varphi_\bfp(\eta)$ is the coefficient of the Fourier 
expansion of the field, i.e., the mode function.
Using the mode function, which is normalized so as to be
$\dot \varphi_\bfp^* \varphi_\bfp-\varphi_\bfp^*\dot \varphi_\bfp=i$,  
the quantized field is constructed as
\begin{eqnarray}
\phi(x)=\sqrt{\frac{\hbar}{L^3}}\sum_{\bfp}
\left(\varphi_\bfp(t)b_\bfp+\varphi^*_{-\bfp}(t) c^\dagger_{-\bfp}\right)
e^{i\bfp\cdot\bfx/\hbar},
\end{eqnarray}
where $L^3$ is the volume of the space, the creation and annihilation 
operators satisfy the commutation relations,
\begin{eqnarray}
[b_\bfp,b^\dagger_{\bfp'}]=\delta_{\bfp,\bfp'}, ~~~~[b_\bfp,b_{\bfp'}]
=[b^\dagger_\bfp,b^\dagger_{\bfp'}]=0,
\end{eqnarray}
and the same relations hold for $c_\bfp$ and $c^\dagger_\bfp$.
We also quantize the free electromagnetic field as,
\begin{eqnarray}
A_\mu=\sqrt{\mu_0\hbar\over L^3}\sum_{\lambda=1,2}\sum_{\bfk}
\epsilon^\lambda_\mu
\left({ e^{-ik\eta}\over \sqrt{2k}} a^\lambda_\bfk+{\rm h.c.}
\right)e^{i\bfk\cdot\bfx},
\end{eqnarray}
where $\epsilon^\lambda_\mu$ denotes the polarization vector, and
$a^{\lambda\dagger}_\bfk$ and $a^\lambda_\bfk$ are the
creation and annihilation operators which satisfy the following 
commutation relation,
\begin{eqnarray}
 [a^\lambda_\bfk,a^{\lambda^\prime\dagger}_{\bfk^\prime}]=\delta^{\lambda\lambda^\prime}\delta_
{\bfk,\bfk^\prime}.
\end{eqnarray}

\begin{figure*}[t]
\includegraphics[scale=1.2]{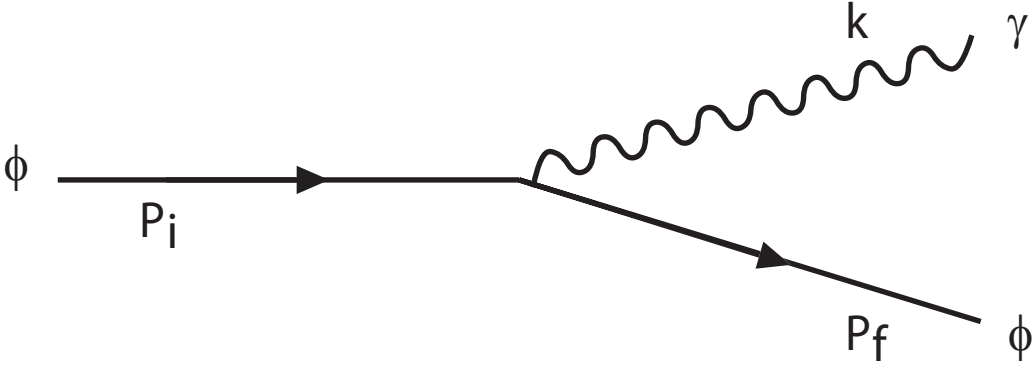}%
\caption{Feynman diagram for the process.
\label{fig1} {}}
\end{figure*}

We consider the process, in which one photon is emitted from 
a charged particle, as shown in Fig.~\ref{fig1}. 
Note that this process is prohibited without the background electric 
field because of the Lorentz invariance of the Minkowski spacetime, 
which ensures existence of the frame that the charged particle
is at rest. However, on the electric field background, we have
the radiation energy from the process, which can be evaluated,
as follows. Using the in-in formalism \cite{Weinberg,Adshead}, 
we may compute the radiation energy at the lowest order of the
coupling constant,
\begin{eqnarray}
E= \sum_{\lambda}\int d^3\bfk\hbar k
 \langle a^{\lambda\dagger}_\bfk a^{\lambda}_\bfk\rangle
=\hbar^{-2}\sum_{\lambda}\int d^3\bfk\hbar k
{\rm Re}
\int^\infty_{-\infty}dt_2\int^{\infty}_{-\infty}
dt_1\left\langle {\rm in}| H_I(t_1)a^{\lambda\dagger}_\bfk a^{\lambda}_{\bfk}H_I(t_2)
|{\rm in}\right\rangle,
\label{Eformula}
\end{eqnarray}
where we adopted the range of the integration from the infinite
past to the infinite future, and 
$|{\rm in}\rangle$ denotes the initial state, which we choose as
one charged particle state with the momentum ${\bf p}_{i}$,
i.e., $|{\rm in}\rangle=b^\dagger_{\bfp_{i}}|0\rangle$, and 
\begin{eqnarray}
H_I(t)&=&-\frac{ie}{\hbar}\int d^3{\bfx}A^\mu
\left\{\left(\partial_\mu- \frac{ie}{\hbar}\bar{A}_\mu\right)\phi^\dagger
\phi
-\phi^\dagger\left(\partial_\mu+\frac{ie}{\hbar}\bar{A}_\mu\right)\phi
\right\}.
\end{eqnarray}
Expression~(\ref{Eformula}) leads to the lowest contribution
corresponding to Fig.~\ref{fig1},
\begin{eqnarray}
&&E=- {e^2\over \epsilon_0} \int {d^3\bfk\over (2\pi)^3}  k\biggl\{
\biggl|\int d\eta {e^{ik\eta}\over {\sqrt{2k}}} \Bigl(
{\partial\over\partial \eta}\varphi_{\bfp_f}(\eta)^* \varphi_{\bfp_{\rm i}}(\eta)-
\varphi_{\bfp_f}^*(\eta){\partial\over\partial \eta}\varphi_{\bfp_{\rm i}}(\eta)
\Bigr)\biggr|^2
\nonumber
\\
&&~~~~~~~~~~~~~~~~~~~~~~~~-\biggl|\int d\eta {e^{ik\eta}\over {\sqrt{2k}}} \Bigl(
{i(\bfp_f-e{\bfA})\over\hbar}\varphi_{\bfp_f}(\eta)^* \varphi_{\bfp_{\rm i}}(\eta)
+\varphi_{\bfp_f}^*(\eta)
{i(\bfp_{\rm i}-e{\bfA})\over\hbar}\varphi_{\bfp_{\rm i}}(\eta)
\Bigr)\biggr|^2
\biggr\},
\end{eqnarray}
where $\bfp_f=\bfp_{\rm i}-\hbar\bfk$, and $\epsilon_0$ is the permittivity
of vacuum, which is related to $\mu_0$ by $\epsilon_0\mu_0=1/c^2=1$. 
Performing the partial integral and using Eq.~(\ref{eqmod}), 
we have
\begin{eqnarray}
&&E=-{e^2\over \epsilon_0} \int {d^3\bfk\over (2\pi)^3}   k\biggl\{
\biggl|\int d\eta  {e^{ik\eta}\over {\sqrt{2k}}}{{\hat \bfk}\cdot(\bfp_{\rm i}+\bfp_f-2e{\bfA})
\over \hbar}\varphi_{\bfp_f}^*(\eta)\varphi_{\bfp_{\rm i}}(\eta)
\biggr|^2
\nonumber
\\
&&~~~~~~~~~~~~~~~~~~~~~~~~-\biggl|\int d\eta {e^{ik\eta}\over {\sqrt{2k}}}
{\bfp_{\rm i}+\bfp_f-2e{\bfA}
\over \hbar}
\varphi_{\bfp_f}^*(\eta)\varphi_{\bfp_{\rm i}}(\eta)
\biggr|^2
\biggr\},
\label{Eformula2}
\end{eqnarray}
where $\hat \bfk$ is the unit vector of $\bfk$, i.e., ${\hat \bfk}=\bfk/|\bfk|$.
We consider the following WKB solution for the mode function
\begin{eqnarray}
\varphi_\bfp(t)={1\over \sqrt{2\Omega_\bfp(t)}} \exp\left[
-i\int^\eta \Omega_\bfp(\eta') d\eta'
\right]
\label{WKBsolution}
\end{eqnarray}
with
\begin{eqnarray}
\Omega_\bfp(\eta)={\sqrt{(\bfp -e{\bfA}(\eta))^2+m^2}/\hbar},
\label{omegabfp}
\end{eqnarray}
then, Eq.~(\ref{Eformula2}) gives
\begin{eqnarray}
E&=& -{e^2\over 2\epsilon_0}{1\over 2^2} \int {d^3\bfk\over (2\pi)^3}   \bigg\{
\biggl|\int d\eta {{\hat \bfk}\cdot(2\bfp_{\rm i}-2e{\bfA}(t)-\hbar\bfk)
\over \hbar \sqrt{\Omega_{\bfp_{\rm i}}(t)}\sqrt{\Omega_{\bfp_f}(t)}}
\exp\Bigl[ik\eta+i\int^\eta(\Omega_{\bfp_f}(t')-\Omega_{\bfp_{\rm i}}(t'))d\eta'\Bigr]
\biggr|^2
\nonumber
\\
&&~~~~~~~~~~~~~~~~~~~~~-\biggl|
\int d\eta {2\bfp_{\rm i}-2e{\bfA}(t)-\hbar\bfk
\over \hbar \sqrt{\Omega_{\bfp_{\rm i}}(t)}\sqrt{\Omega_{\bfp_f}(t)}}
\exp\Bigl[ik\eta+i\int^\eta(\Omega_{\bfp_f}(t')-\Omega_{\bfp_{\rm i}}(t'))d\eta'\Bigr]
\biggr|^2
\bigg\}
\end{eqnarray}
with
\begin{eqnarray}
&&\Omega_{\bfp_{\rm i}}(\eta)={1\over \hbar}\sqrt{(\bfp_{\rm i} - e{\bfA}(t))^2+m^2},
~~~~~~~~
\Omega_{\bfp_f}(\eta)={1\over \hbar}\sqrt{(\bfp_{\rm i}-\hbar\bfk - e{\bfA}(t))^2+m^2}.
\end{eqnarray}
%
%
In order to evaluate the quantum correction,
we consider the expansion in terms of a power series of $\hbar$. 
Up to the order of  ${\cal O}(\hbar)$, we have 
\begin{eqnarray*}
&&\Omega_{\bfp_f}-\Omega_{\bfp_{\rm i}}
\simeq-{\bfk\cdot(\bfp_{\rm i}-e{\bfA})\over\sqrt{(\bfp_{\rm i}-e{\bfA})^2+m^2}}+{\hbar \over 2} \biggl(
{k^2\over\sqrt{(\bfp_{\rm i}-e{\bfA})^2+m^2} }-
{(\bfk\cdot(\bfp_{\rm i}-e{\bfA}))^2\over
\sqrt{(\bfp_{\rm i}-e{\bfA})^2+m^2}^3}\biggr), 
\\
&&{1\over \hbar \sqrt{\Omega_{\bfp_{\rm i}}}\sqrt{\Omega_{\bfp_f}}}
\simeq{1\over \sqrt{(\bfp_{\rm i}-e{\bfA})^2+m^2}}\biggl(
1+{\hbar\over 2}{\bfk\cdot(\bfp_{\rm i}-e{\bfA})\over [(\bfp_{\rm i}-e{\bfA})^2+m^2]}
\biggr),
\end{eqnarray*}
then
\begin{eqnarray}
E
&=&- 
{e^2\over 2\epsilon_0}
 \int {d^3 \bfk\over (2\pi)^3}   \int d\xi \int d\xi'
\left\{\Bigl({{\hat \bfk}\cdot{d\bfx\over d\xi}}\Bigr)
       \Bigl({{\hat \bfk}\cdot{d\bfx'\over d\xi'}}\Bigr)
-{d\bfx\over d\xi}\cdot{d\bfx'\over d\xi'}
\right\}e^{ik\xi-ik\xi'}
\nonumber
\\
&&\times\biggl[
1+{\hbar k\over 2}{\hat\bfk\cdot(\bfp_{\rm i}-e{\bfA})\over (\bfp_{\rm i}-e\bfA)^2+m^2}
+{\hbar k\over 2}{\hat\bfk\cdot(\bfp_{\rm i}-e{\bfA}')\over (\bfp_{\rm i}-e\bfA')^2+m^2}
\nonumber
\\
&&~~~~+{i\hbar\over 2}\int_{\eta'}^\eta
\biggl({k^2\over \sqrt{(\bfp_{\rm i}-e\bfA'')^2+m^2}}
-{k^2(\hat\bfk\cdot(\bfp_{\rm i}-e\bfA''))^2\over \sqrt{(\bfp_{\rm i}-e\bfA'')^2+m^2}^3}
\biggr)d\eta''
+{\cal O}(\hbar^2)
\biggr],
\label{Eexpression}
\end{eqnarray}
where we used the notations $\bfx=\bfx(t)$, $\bfx'=\bfx(t')$, 
$\bfA=\bfA(\eta),~\bfA'=\bfA(\eta'),~\bfA''=\bfA(\eta'')$,
and we introduced the new variable $\xi$ instead of $\eta$
\begin{eqnarray}
\xi=\eta-\int^\eta {{\hat \bfk}\cdot(\bfp_{\rm i}-e{\bfA}(t''))
\over \sqrt{(\bfp_{\rm i}-e{\bfA}(t''))^2+m^2}}d\eta'', 
\label{dexfi}
\end{eqnarray}
and $\xi'$ is defined in the same way as $\xi$ but with replacing $t$ by $t'$.
Furthermore, we introduce the quantities parametrized by $\eta$ (or
$\xi$),
\begin{eqnarray}
&&{d\bfx\over d\tau}=\bfp_{\rm i}-e{\bfA},\\
&&{d\eta\over d\tau}=\sqrt{(\bfp_{\rm i}-e{\bfA})^2+m^2},
\end{eqnarray}
then Eq.~(\ref{Eexpression}) is rephrased as
\begin{eqnarray}
E&=&-
{e^2\over 2\epsilon_0}
\int {d^3\bfk\over (2\pi)^3}   \int d\xi \int d\xi'
\left\{\Bigl({{\hat \bfk}\cdot{d\bfx\over d\xi}}\Bigr)
       \Bigl({{\hat \bfk}\cdot{d\bfx'\over d\xi'}}\Bigr)
-{d\bfx\over d\xi}\cdot{d\bfx'\over d\xi'}
\right\}e^{ik\xi-ik\xi'}
\nonumber
\\
&&\times\biggl[
1+{\hbar k\over 2}\biggl(
\hat\bfk\cdot{d\bfx\over d\eta} {d\tau\over d\eta}+
\hat\bfk\cdot{d\bfx'\over d\eta'} {d\tau'\over d\eta'}
\biggr)
+{i\hbar k^2\over 2}\int_{\eta'}^\eta d\tau''
\biggl(1-\biggl(\hat\bfk\cdot{d\bfx''\over d\eta''}\biggr)^2
\biggr)\biggr]+{\cal O}(\hbar^2).
\label{EexpressionII}
\end{eqnarray}
The mathematical technique adopted in Ref.~\cite{HW} is 
equivalent to replacing $k$ in Eq.~(\ref{EexpressionII})
by the partial differentiation with respect to $\xi$ or $\xi'$ which 
operates to $e^{ik\xi-ik\xi'}$.
The partial integrations lead to
\begin{eqnarray}
E=E^{(0)}+E^{(1)}+{\cal O}(\hbar^2),
\end{eqnarray}
where we defined
\begin{eqnarray}
E^{(0)}&=&-
{e^2\over 2 \epsilon_0(2\pi)^3}
\int d\Omega_{\hat\bfk} \int_0^\infty dk  \int d\xi \int d\xi' 
e^{ik(\xi-\xi')}
\biggl(\Bigl({{\hat \bfk}\cdot{d^2\bfx\over d\xi^2}}\Bigr)
       \Bigl({{\hat \bfk}\cdot{d^2\bfx'\over d\xi'^2}}\Bigr)
-{d^2\bfx\over d\xi^2}\cdot{d^2\bfx'\over d\xi'^2}
\biggr),
\label{E0}
\\
E^{(1)}&=&-
{e^2\over 2 \epsilon_0(2\pi)^3}
\int d\Omega_{\hat\bfk} \int_0^\infty dk  \int d\xi \int d\xi' 
e^{ik(\xi-\xi')}
\nonumber
\\
&&\times\Biggl\{{i\hbar\over 4}\biggl({d\over d\xi}-{d\over d\xi'}\biggr)
{d\over d\xi}{d\over d\xi'}\biggl[\biggl(\Bigl({{\hat \bfk}\cdot{d\bfx\over d\xi}}\Bigr)
       \Bigl({{\hat \bfk}\cdot{d\bfx'\over d\xi'}}\Bigr)
-{d\bfx\over d\xi}\cdot{d\bfx'\over d\xi'}
\biggr)\biggl(
\hat\bfk\cdot{d\bfx\over d\eta} {d\tau\over d\eta}+
\hat\bfk\cdot{d\bfx'\over d\eta'} {d\tau'\over d\eta'}
\biggr)\biggr]
\nonumber
\\
&&+
{i\hbar \over 2}
{d^2\over d\xi^2}{d^2\over d\xi'^2}\biggl[\biggl(\Bigl({{\hat \bfk}\cdot{d\bfx\over d\xi}}\Bigr)
       \Bigl({{\hat \bfk}\cdot{d\bfx'\over d\xi'}}\Bigr)
-{d\bfx\over d\xi}\cdot{d\bfx'\over d\xi'}
\biggr)
\int_{\xi'(\eta')}^{\xi(\eta)} d\xi''{d\tau''\over d\xi''}
\biggl(1-\biggl(\hat\bfk\cdot{d\bfx''\over d\eta''}\biggr)^2
\biggr)\biggr]\Biggl\},
\label{E1}
\end{eqnarray}
where $E^{(0)}$ and $E^{(1)}$ are the terms of the order of
$\hbar^0$ and $\hbar^1$, respectively. Here, we assumed the 
boundary terms can be neglected, as is the case in Ref.~\cite{HW}. 
The integration with respect to $k$ yields
\begin{eqnarray}
E^{(0)}&=&
{e^2\over (4\pi)^2 \epsilon_0}
\int d\Omega_{\hat\bfk} \int d\xi 
\biggl(
\left({d^2\bfx\over d\xi^2}\right)^2-
\left({{\hat \bfk}\cdot{d^2\bfx\over d\xi^2}}\right)^2
\biggr).
\label{Eee0}
\end{eqnarray}
The expression (\ref{Eee0}) yields the classical formula of the Larmor 
radiation from a charged particle.
The first-order quantum correction of the order of $\hbar$ is described 
by Eq.~(\ref{E1}), which yields
\begin{eqnarray}
E^{(1)}&=&
{e^2\hbar \over  (4\pi)^3\epsilon_0}\int d\Omega_{\hat\bfk} 
\int d\xi \int d\xi' {1\over \xi-\xi'}
\nonumber
\\
&&\times\Biggl\{\biggl({d\over d\xi}-{d\over d\xi'}\biggr)
{d\over d\xi}{d\over d\xi'}\biggl[\biggl(\Bigl({{\hat \bfk}\cdot{d\bfx\over d\xi}}\Bigr)
       \Bigl({{\hat \bfk}\cdot{d\bfx'\over d\xi'}}\Bigr)
-{d\bfx\over d\xi}\cdot{d\bfx'\over d\xi'}
\biggr)\biggl(
\hat\bfk\cdot{d\bfx\over d\eta} {d\tau\over d\eta}+
\hat\bfk\cdot{d\bfx'\over d\eta'} {d\tau'\over d\eta'}
\biggr)\biggr]
\nonumber
\\
&&+2
{d^2\over d\xi^2}{d^2\over d\xi'^2}\biggl[\biggl(\Bigl({{\hat \bfk}\cdot{d\bfx\over d\xi}}\Bigr)
       \Bigl({{\hat \bfk}\cdot{d\bfx'\over d\xi'}}\Bigr)
-{d\bfx\over d\xi}\cdot{d\bfx'\over d\xi'}
\biggr)
\int_{\xi'(\eta')}^{\xi(\eta)} d\xi''{d\tau''\over d\xi''}
\biggl(1-\biggl(\hat\bfk\cdot{d\bfx''\over d\eta''}\biggr)^2
\biggr)\biggr]\Biggl\}.
\label{E1b}
\end{eqnarray}
Equation~(\ref{E1b}) transforms into Eq.~(\ref{Eeee1}). Other useful
formulas are summarized in the appendix A.

\section{Approximate formulas}
In the non-relativistic limit, where the velocity ${\bf v}={d\bf x}/d\eta$
is small enough compared with the velocity of light, $|\bfv|\ll 1$, 
Eqs.~(\ref{Eee0}) and (\ref{E1b}) reduce to 
\begin{eqnarray}
&&E^{(0)}={e^2\over 6\pi\epsilon_0}\int dt \dot\bfv(t)\cdot\dot\bfv(t),
\label{EE0}
\\
&&E^{(1)}={e^2\hbar\over 6\pi^2\epsilon_0m} \int dt\int dt'
{\ddot\bfv(t) \cdot\dot \bfv(t')-\dot\bfv(t)\cdot\ddot \bfv(t')\over t-t'},
\label{EE1}
\end{eqnarray}
respectively. A brief derivation is summarized in the appendix A.
Equation~(\ref{EE1}) was found for the first time by 
Higuchi and Walker in Ref.~\cite{HW}. 
In the case of the periodic electric field, $|{\bf E}|=E_0\sin\omega t$,
where $E_0$ is a constant,
we have the periodic acceleration, $|\dot \bfv|=(eE_0/m)\sin\omega t$.
Then
\begin{eqnarray}
&&{dE^{(0)}\over dt}= {e^4E_0^2\over m^2}{\sin^2\omega t\over 6\pi\epsilon_0},
\\
&&{dE^{(1)}\over dt}=-{\hbar e^4E_0^2\over m^2}{\omega \over  12\pi\epsilon_0m}.
\label{defasdfa}
\end{eqnarray}
After taking an average over a long time-duration, we have
\begin{eqnarray}
{E^{(1)}\over E^{(0)}}=-{\hbar \omega \over mc^2},
\end{eqnarray}
where $c$ is the light velocity, which is restored here.
The quantum effect becomes important when the time scale of the
acceleration multiplied by $c$ is comparable to the Compton 
wavelength, namely, when the wavelike feature of the particle appears.

\begin{figure*}[t]
\includegraphics[scale=0.6]{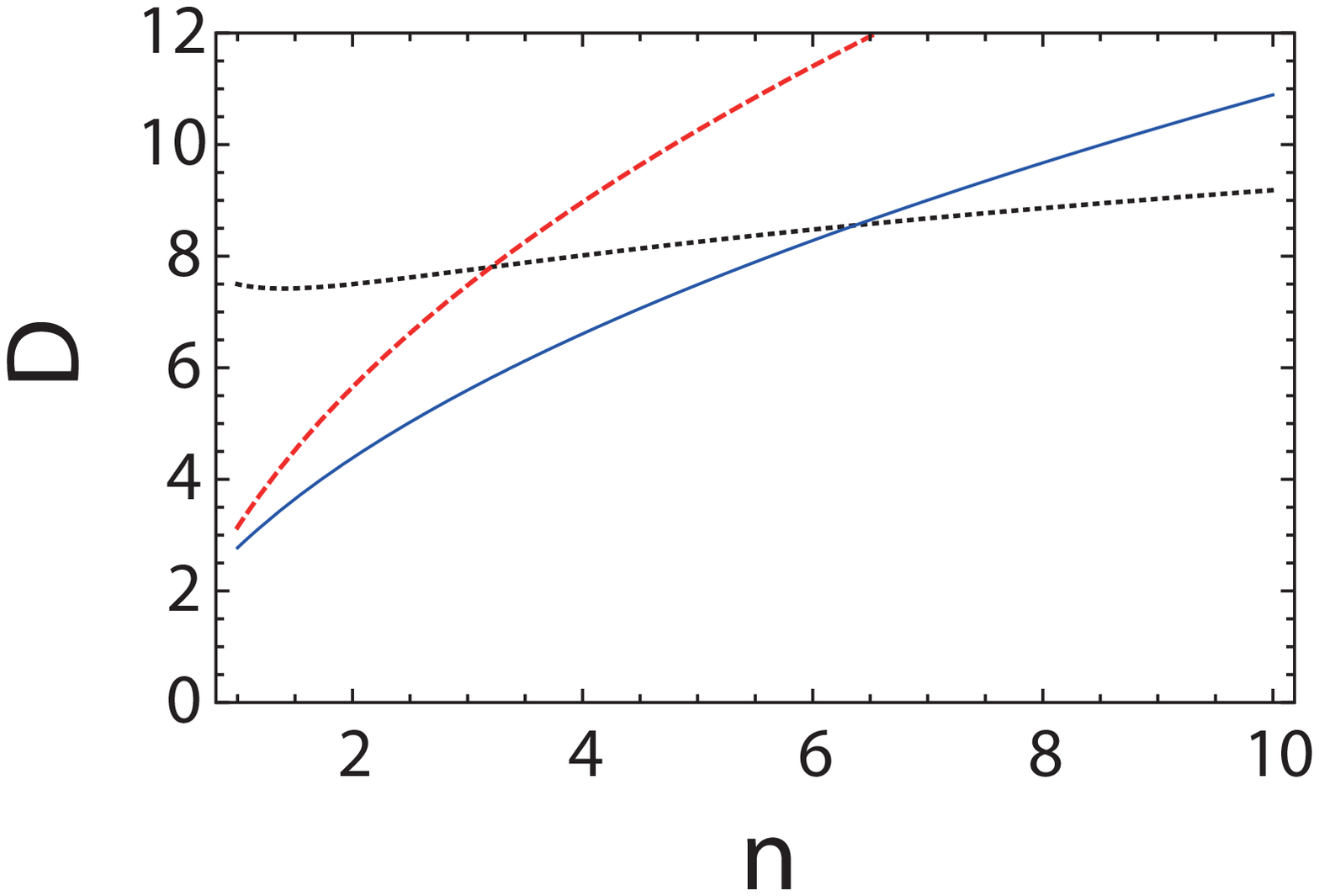}%
\caption{$D$ as a function of $n$ for cases (1) -- (3).
The dotted curve is case (1) $f(\tau)=1-(\tau^2)^n$ 
for $-1\leq \tau\leq 1$ and $f(\tau)=0$ for $|\tau|>1$, 
the (red) dashed curve is case (2)
$f(\tau)=1/(1+\tau^2)^n$, and 
the (blue) solid curve is case (3)
$f(\tau)=1/(\cosh\tau)^n$.
\label{fig2} {}}
\includegraphics[scale=0.6]{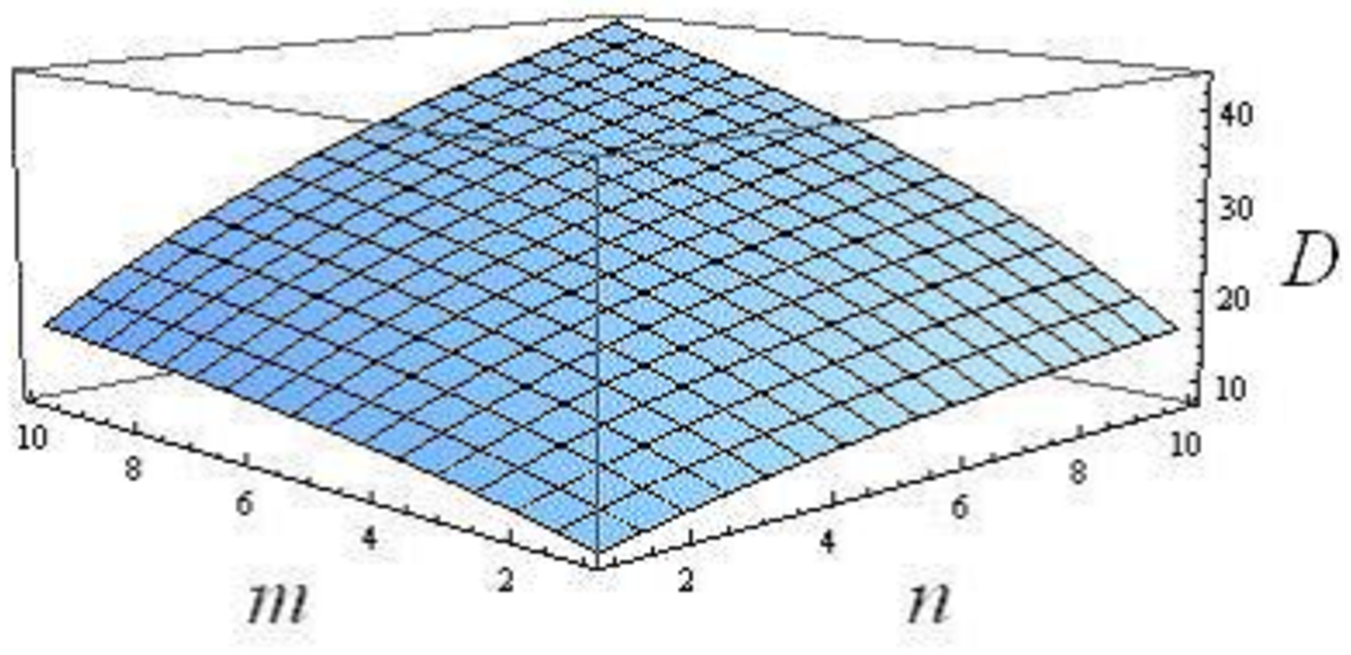}%
\caption{$D$ as a function of $m$ and $n$ for model (4).
\label{fig3} {}}
\end{figure*}

Let us consider a more general case, when the electric field
$|{\bf E}|=E_0f(t/t_0)$, where $f$ is a function of $t/t_0$
with a constant $t_0(>0)$. In this case, the acceleration is
$|\dot{\bf v}|=(eE_0/m)f(t/t_0)$, and
Eqs.~(\ref{Eee0}) and (\ref{E1b})
give
\begin{eqnarray}
&& E^{(0)}=\frac{e^4E_0^2 t_0}{6\pi\epsilon_0m^2}
\int d\tau f^2(\tau)
,\\
&& E^{(1)}=-\frac{\hbar e^4E_0^2 }{6\pi^2\epsilon_0m^3}
\int\int d\tau d\tau' 
{f(\tau)f_{,\tau'}(\tau')-f(\tau')f_{,\tau}(\tau)\over \tau-\tau'},
\end{eqnarray}
then, we have
\begin{eqnarray}
&&\frac{E^{(1)}}{E^{(0)}}=-\frac{\hbar}{\pi mc^2t_0}D,
\end{eqnarray}
where we defined 
\begin{eqnarray}
D=
\left({\int d\tau'' f^2(\tau'')}\right)^{-1}
\displaystyle{{
\int\int d\tau d\tau' 
{f(\tau)f_{,\tau'}(\tau')-f(\tau')f_{,\tau}(\tau)\over \tau-\tau'}
}},
\label{defD}
\end{eqnarray}
where $f_{,\tau}(\tau)$ means the differentiation of $f(\tau)$ 
with respect to $\tau$.
Here, let us consider the following three cases: 
(1) $f(\tau)=1-\left(\tau^2\right)^{n}$
for $|\tau|\leq 1$ and $f(\tau)=0$ for $|\tau|> 1$,
(2) $f(\tau)=1/(1+\tau^2)^n$ for $-\infty<\tau< \infty$, 
(3) $f(\tau)=1/(\cosh\tau)^n$ for $-\infty<\tau< \infty$, 
and (4) $f(\tau)=\left(1-|\tau|\right)^{n}$
for $-1\leq\tau\leq 0$, $f(\tau)=\left(1-|\tau|\right)^{m}$
for $0<\tau\leq 1$,  and $f(\tau)=0$ for $|\tau|> 1$. 
Figure \ref{fig2} show $D$ as a function of $n$ for cases
(1) -- (3), for which one can see that $D$ is positive.
Figure \ref{fig3} show $D$ as a function of $n$ and $m$ for case (4).

Thus, in all of the above cases, the first-order quantum correction 
$E^{(1)}$ is negative. Also, the quantum effect is very small 
as long as the motion of the particle is non-relativistic. 
This result is consistent with that found in Refs.~\cite{HW,NSY}.
The quantum effect might become important when the emitted photon energy
becomes of order of $mc^2$~\cite{HW}. Note that this speculation is based on 
the result with the non-relativistic approximation.

Next, let us consider the relativistic limit, $|\bfp_{\rm i}|\gg |e{\bf A}|,~m$.
For simplicity, we consider the case when the direction of the particle 
motion is always parallel to that of the background electric field, i.e., 
$\bfv \propto \bfA$.
Namely, we consider the case when the directions of the particle's 
motion and 
the background electric field are parallel at any moment, and 
adopt this direction as the $z$ axis. 
Then, we may write $\bfA=(0,0,A(t))$, $\dot \bfA=(0,0,-E(t))$, 
$\bfv=(0,0,v)$, and $\bfp_{\rm i}=(0,0,p_{\rm i})$. 
In this case, we have
\begin{eqnarray}
&&E^{(0)}={e^2\over (4\pi)^2\epsilon_0}\int d\Omega_{\hat\bfk} (1-\cos^2\theta)\int dt   {m^4\over p_{\rm i}^6}{e^2\dot A^2(t)\over (1-v\cos\theta)^5}.
\label{ER0}
\end{eqnarray}
The integration with respect to ${\hat\bfk}$ yields
\begin{eqnarray}
&&E^{(0)}={1\over 6\pi\epsilon_0}{m^4e^4\over p_{\rm i}^6}\int dt  {\dot A^2(t)\over 
(1-v^2)^3}.
\label{ER0b}
\end{eqnarray}
We consider the case $p_i\gg |eA|,~m$. We also assume 
$|A|\sim |\dot A/\omega| \sim
|\ddot A/\omega^2|$, where $1/\omega$ is a timescale of 
a time-varying background electric field.
In this relativistic limit, we have the leading order 
expression for the quantum correction (see also appendix A), 
\begin{eqnarray}
E^{(1)}&\simeq&-
{e^2\hbar\over  (4\pi)^3\epsilon_0}\int d\Omega_{\hat\bfk} (1-\cos^2\theta)
\int d\xi \int d\xi' {1\over \xi-\xi'}
{m^2\over p_{\rm i}^5}{1\over (1-v\cos\theta)^2(1-v'\cos\theta)^2}
\nonumber
\\
&&~~~~~\times e^2\Bigg\{\ddot A(t) \dot A(t')\left(
{-v^2\cos\theta\over (1-v\cos\theta)(1-v'\cos\theta)}
+{(2+v'\cos\theta)v'\over (1-v\cos\theta)^2}
\right)
\nonumber
\\&&~~~~~~~~~~-\ddot A(t') \dot A(t)\left(
{-v'^2\cos\theta\over (1-v\cos\theta)(1-v'\cos\theta)}
+{(2+v\cos\theta)v\over (1-v'\cos\theta)^2}
\right)
\Bigg\}.
\label{E1D}
\end{eqnarray}
Adopting the approximation, $v=v'=\bar v\simeq 1$, and $\xi-\xi'\simeq
(t-t')(1-{\bar v}\cos\theta)$, we have
\begin{eqnarray}
&&E^{(1)}\simeq-{e^2\hbar\over 4(2\pi)^3\epsilon_0} \int d\Omega_{\hat\bfk} (1-\cos^2\theta) \int dt\int dt' {1\over (1-\bar v\cos\theta)^5} {m^2\over p_{\rm i}^5}
{e^2(\ddot A(t) \dot A(t')-  \dot A(t)\ddot A(t')) \over t-t'}.
\label{ER1}
\end{eqnarray}
The integration with respect to ${\hat\bfk}$ yields
\begin{eqnarray}
&&E^{(1)}\simeq-{e^4\hbar\over 3(2\pi)^2\epsilon_0} {m^2\over p_{\rm i}^5}\int dt\int dt' 
{1\over (1-\bar v^2)^3} 
{\ddot A(t) \dot A(t')-  \dot A(t)\ddot A(t') \over t-t'}.
\label{ER11}
\end{eqnarray}

In the case of the periodic background of the electric field, 
$\dot A(t)  = -E_0\sin\omega t$, where $E_0$ is a constant, 
we have
\begin{eqnarray}
&&{dE^{(0)}\over dt}={e^4 m^4\over 6\pi \epsilon_0 p_{\rm i}^6}{E_0^2\cos^2 \omega t\over (1-v^2)^3},
\\
&&{dE^{(1)}\over dt}={\hbar e^4 m^2\over 12\pi \epsilon_0 p_{\rm i}^5}{E_0^2\omega \over (1-v^2)^3}.
\end{eqnarray}
After averaging over sufficiently long time-duration, we have
\begin{eqnarray}
{E^{(1)}\over E^{(0)}}={p_{\rm i}\over mc}{ \hbar \omega \over mc^2}.
\label{relarela}
\end{eqnarray}
Note that the quantum correction $E^{(1)}$ is positive, 
which is a contrast to the non-relativistic 
case.

Similar to the non-relativistic limit, 
we next consider the case, $\dot{A}(t)=-E_0f\left({t}/{t_0}\right)$, 
with a general function $f(\tau)$. 
In the case, we have
\begin{eqnarray}
&& E^{(0)}=\frac{e^4m^4E_0^2t_0}{6\pi\epsilon_0 p_{\rm i}^6(1-\bar{v})^3}
\int d\tau f^2(\tau)
,\\
&& E^{(1)}=\frac{\hbar e^4m^2E_0^2}{12\pi^2\epsilon_0p_{\rm i}^5(1-\bar{v})^3}
\int\int d\tau d\tau' 
{f(\tau)f_{,\tau'}(\tau')-f(\tau')f_{,\tau}(\tau)\over \tau-\tau'},
\end{eqnarray}
and
\begin{eqnarray}
\frac{E^{(1)}}{E^{(0)}}=\frac{p_{\rm i}}{\pi mc}\frac{\hbar}{mc^2t_0}D,
\end{eqnarray}
where $D$ is defined by Eq.~(\ref{defD}). When we adopt the 
three function of $f(\tau)$ of (1) -- (3) as in the case of the 
non-relativistic limit, $D$ is positive. Thus, 
in contrast to the non-relativistic case, the quantum correction 
$E^{(1)}$ is positive again, for all the cases in the present paper. 

For the radiation from an electron in a periodic electric field, e.g., 
by a laser field, Eq.~(\ref{relarela}) is estimated as 
\begin{eqnarray}
 {E^{(1)}\over E^{(0)}}\sim 2.6\times10^{-3}
\left(\frac{p_{\rm i}c}{{\rm GeV}}\right)
\left(\frac{mc^2}{0.5 {\rm MeV}}\right)^{-2}
\left(\frac{\omega}{10^{15} s^{-1}}\right),
\label{180037_3Nov10}
\end{eqnarray}
where $\omega \sim 10^{15}s^{-1}$ corresponds to an x-ray laser.
The quantum effect becomes significant when the electron kinetic 
energy reaches the TeV scale.
The above formula is derived under the condition, $p_i\gg |eA|,~m$. 
For a periodic electric field of large amplitude, $p_i\sim |eA|$, 
the condition of the relativistic motion cannot 
be always guaranteed, because the physical momentum 
might become $|{\bf p}_i-e{\bf A}|\sim m$.
In this case, it is difficult to express the quantum correction 
in a simple analytic form. We need a more general 
treatment including fully numerical calculation, 
because the non-locality plays an important role.
Potentially, there is a lot of room for discussion about how to detect 
the quantum effect of the Larmor radiation experimentally, but this 
is outside of the scope of the present paper.

\section{Summary}
In the present paper, we obtained the general formula, Eq.~(\ref{E1b})
or Eq.~(\ref{Eeee1}), for 
the first-order quantum correction to the Larmor radiation from 
a charged particle moving in spatially homogeneous time-dependent 
electric field. 
This formula reproduces the same result as that in Ref.~\cite{HW},
in the limit of a non-relativistic motion of the charged particle. 
Our result is useful to investigate the case of a relativistic motion.
When the direction of a particle's motion is parallel
to that of the background electric field, a simple formula was derived.
In the limit of the relativistic motion of the charged particle, 
we obtained the formula (\ref{E1D}). 
Similar to the case of the non-relativistic motion \cite{HW}, 
the leading quantum effect is described by a non-local difference 
between $\dot{A}(t)\ddot{A}(t')$ and $\ddot{A}(t)\dot{A}(t')$, 
as is demonstrated in Eq.~(\ref{ER11}).
This quantum effect disappears when $\dot A$ is constant.
Note that Eq.~(\ref{ER11}) is the leading term in the limit of
the ultra-relativistic motion, assuming $p_i\gg eA,~m$ and $|A|\sim
|\dot A/\omega|\sim |\ddot A/\omega^2|$.
We discarded the other sub-leading terms. For example, the term
in proportion to $\dot{A}(t)\dot{A}(t')(\dot{A}(t)-\dot{A}(t'))$
appears in the sub-leading terms, but also disappears when $\dot A$ 
is constant.
Thus, the essence of the quantum effect of the Larmor radiation
should be the nonlocality, which reflects the fact that the exact 
solution of motion cannot be represented with simple classical 
trajectories in quantum theory \cite{HW}.

We also note that the expression in the non-relativistic limit 
(\ref{EE1}) is not simply connected to that in the relativistic 
limit (\ref{ER11}). 
The leading contributions for these opposite limits come
from different sources. 
In the non-relativistic limit, the leading contribution 
comes only from terms in (\ref{A2}), i.e., 
the phase of the mode function. 
On the other hand, in the relativistic limit, 
the leading contribution comes from both (\ref{A1}) and (\ref{A2}), 
i.e., the amplitude and the phase of the mode function.
An interesting question might be how these facts are related to
the difference of our final results in the opposite limits.

By applying the formula to the cases of a periodic 
acceleration and possible function of acceleration, 
it was demonstrated that the leading quantum effect enhances 
the radiation in the relativistic limit and that it 
decreases in the non-relativistic limit. 
This quantum effect will become important when 
the incident kinetic electron energy approaches 
${\rm TeV}$ scale for a periodic electric 
field background with an x-ray laser.
However, this result is obtained assuming that the charged particle 
is moving in the direction parallel to that of the background electric 
field. In a practical situation, this assumption is somewhat ideal.
Here too there is a lot of room for further investigations of
more general cases (cf. Ref.\cite{HW}), but this is outside of the 
scope of the present paper.  

Our work, which is based on the QED theoretical framework,   
will be useful to investigate the feature of the radiation 
from an electron under a strong electric field.
Investigation of the quantum effect in the Larmor radiation 
could be related to the subject of testing the QED process 
in the strong field background. 
For example, Chen and Tajima claimed the possibility of
detecting the Unruh effect in the radiation from 
an electron under an ultraintense laser background \cite{ChenTajima}. 
Possible signature of the Unruh effect in the radiation from 
an electron accelerated by an electric field of strong lasers 
is under debate (cf.~\cite{Schutzhold,HiguchiUnruh,Iso}).
According to Ref.~\cite{ChenTajima}, the radiation from the 
Unruh effect could be of order of $\hbar$. 
The characteristic signature of the Unruh effect claimed in 
\cite{ChenTajima} is in proportion to $E_0^3$ at order of 
$\hbar$. As mentioned in the above, in our approach, 
the term in proportion to 
$\dot{A}(t)\dot{A}(t')(\dot{A}(t)-\dot{A}(t'))$ appears 
in the sub-leading terms in evaluating Eq.~(\ref{E1b}).
This might give a contribution in proportion to $E_0^3$.
However, the angular dependence is different.
In our approach, the quantum radiation of the order $\hbar$ 
emitted in the direction of the motion, $\theta=0$, is exactly 
zero from Eq.~(\ref{EexpressionII}). This is a
difference between our result and the prediction in 
Ref.~\cite{ChenTajima}, which might be tested experimentally.
However, in our approach, it is difficult to separate the 
signature of the Unruh effect from other effects, 
even if they existed.  This is a disadvantage of our approach.

In Ref.~\cite{NSY}, the quantum radiation from a charged particle 
moving in an expanding or contracting universe was investigated. 
It was shown that the radiation can be regarded as the Larmor 
radiation from a charged particle in an decelerated 
(accelerated) motion, because the physical momentum of the 
particle decreases (increases)  as the background universe 
expands (contracts) \cite{NSY,HiguchiUniv}. 
The approach developed
in the present paper is useful to investigate the 
quantum effect of this process \cite{cfu2}.



\vspace{0.2cm}
{\it Acknowledgment}
K.Y. thanks K.~Homma, T.~Takahashi, H.~Nomura, M.~Sasaki, H.~Okamoto, 
K.~Yokoya and A.~Higuchi for useful communication when 
the topic of the present paper was intiated.
We thank R. Kimura for useful discussions and comments.   
This work was supported by the Japan Society for Promotion
of Science (JSPS) Grants-in-Aid
for Scientific Research (No.~21540270,No.~21244033).
This work was also supported by JSPS Core-to-Core Program 
``International Research Network for Dark Energy.''

\begin{appendix}
\section{Brief Summary of Derivation of Approximate Formulas}
It is straightforward to derive the following formulas,
\begin{eqnarray}\label{A1}
&&\biggl({d\over d\xi}-{d\over d\xi'}\biggr)
{d\over d\xi}{d\over d\xi'}\biggl[\biggl(\Bigl({{\hat \bfk}\cdot{d\bfx\over d\xi}}\Bigr)
       \Bigl({{\hat \bfk}\cdot{d\bfx'\over d\xi'}}\Bigr)
-{d\bfx\over d\xi}\cdot{d\bfx'\over d\xi'}
\biggr)\biggl(
\hat\bfk\cdot{d\bfx\over d\eta} {d\tau\over d\eta}+
\hat\bfk\cdot{d\bfx'\over d\eta'} {d\tau'\over d\eta'}
\biggr)\biggr]
\nonumber
\\
&&~~~~=\biggl(\Bigl({{\hat \bfk}\cdot{d^3\bfx\over d\xi^3}}\Bigr)
          \Bigl({{\hat \bfk}\cdot{d^2\bfx'\over d\xi'^2}}\Bigr)
          -{d^3\bfx\over d\xi^3}\cdot{d^2\bfx'\over d\xi'^2}
         -\Bigl({{\hat \bfk}\cdot{d^2\bfx\over d\xi^2}}\Bigr)
          \Bigl({{\hat \bfk}\cdot{d^3\bfx'\over d\xi'^3}}\Bigr)
          +{d^2\bfx\over d\xi^2}\cdot{d^3\bfx'\over d\xi'^3}\biggr)
\nonumber\\&&~~~~~~\times
   \biggl(\hat\bfk\cdot{d\bfx\over d\eta} {d\tau\over d\eta}+
          \hat\bfk\cdot{d\bfx'\over d\eta'} {d\tau'\over d\eta'}\biggr)
\nonumber\\
&&~~~~+2\biggl(\Bigl({{\hat \bfk}\cdot{d^2\bfx\over d\xi^2}}\Bigr)
          \Bigl({{\hat \bfk}\cdot{d^2\bfx'\over d\xi'^2}}\Bigr)
          -{d^2\bfx\over d\xi^2}\cdot{d^2\bfx'\over d\xi'^2}\biggr)
   \biggl(
   {d\over d\xi}\Bigl(\hat\bfk\cdot{d\bfx\over d\eta} {d\tau\over d\eta}\Bigr)
-  {d\over d\xi'}\Bigl(\hat\bfk\cdot{d\bfx'\over d\eta'} {d\tau'\over d\eta'}\Bigr)
  \biggr)
\nonumber\\
&&~~~~+\biggl(\Bigl({{\hat \bfk}\cdot{d\bfx\over d\xi}}\Bigr)
          \Bigl({{\hat \bfk}\cdot{d^2\bfx'\over d\xi'^2}}\Bigr)
          -{d\bfx\over d\xi}\cdot{d^2\bfx'\over d\xi'^2}\biggr)
{d^2\over d\xi^2}\Bigl(\hat\bfk\cdot{d\bfx\over d\eta} {d\tau\over d\eta}\Bigr)
\nonumber\\
&&~~~~-\biggl(\Bigl({{\hat \bfk}\cdot{d^2\bfx\over d\xi^2}}\Bigr)
          \Bigl({{\hat \bfk}\cdot{d\bfx'\over d\xi'}}\Bigr)
          -{d^2\bfx\over d\xi^2}\cdot{d\bfx'\over d\xi'}\biggr)
{d^2\over d\xi'^2}\Bigl(\hat\bfk\cdot{d\bfx'\over d\eta'} {d\tau'\over d\eta'}\Bigr)
\nonumber\\
&&~~~~+\biggl(\Bigl({{\hat \bfk}\cdot{d^3\bfx\over d\xi^3}}\Bigr)
          \Bigl({{\hat \bfk}\cdot{d\bfx'\over d\xi'}}\Bigr)
          -{d^3\bfx\over d\xi^3}\cdot{d\bfx'\over d\xi'}\biggr)
{d\over d\xi'}\Bigl(\hat\bfk\cdot{d\bfx'\over d\eta'} {d\tau'\over d\eta'}\Bigr)
\nonumber\\
&&~~~~-\biggl(\Bigl({{\hat \bfk}\cdot{d\bfx\over d\xi}}\Bigr)
          \Bigl({{\hat \bfk}\cdot{d^3\bfx'\over d\xi'^3}}\Bigr)
          -{d\bfx\over d\xi}\cdot{d^3\bfx'\over d\xi'^3}\biggr)
{d\over d\xi}\Bigl(\hat\bfk\cdot{d\bfx\over d\eta} {d\tau\over d\eta}\Bigr)
\end{eqnarray}
and
\begin{eqnarray}\label{A2}
&&{d^2\over d\xi^2}{d^2\over d\xi'^2}\biggl[\biggl(\Bigl({{\hat \bfk}\cdot{d\bfx\over d\xi}}\Bigr)
       \Bigl({{\hat \bfk}\cdot{d\bfx'\over d\xi'}}\Bigr)
-{d\bfx\over d\xi}\cdot{d\bfx'\over d\xi'}
\biggr)
\int_{\xi'(\eta')}^{\xi(\eta)} d\xi''{d\tau''\over d\xi''}
\biggl(1-\biggl(\hat\bfk\cdot{d\bfx''\over d\eta''}\biggr)^2
\biggr)\biggr]
\nonumber\\
&&~~~~=\biggl(\Bigl({{\hat \bfk}\cdot{d^3\bfx\over d\xi^3}}\Bigr)
          \Bigl({{\hat \bfk}\cdot{d^3\bfx'\over d\xi'^3}}\Bigr)
          -{d^3\bfx\over d\xi^3}\cdot{d^3\bfx'\over d\xi'^3}\biggr)
   \int_{\xi'}^\xi d\xi'' {d\tau''\over d\xi''}
   \biggl(1-\Bigl(\hat\bfk\cdot{d\bfx''\over d\eta''}\Bigr)^2\biggr)
\nonumber\\
&&~~~~~~-2\biggl(\Bigl({{\hat \bfk}\cdot{d^3\bfx\over d\xi^3}}\Bigr)
          \Bigl({{\hat \bfk}\cdot{d^2\bfx'\over d\xi'^2}}\Bigr)
          -{d^3\bfx\over d\xi^3}\cdot{d^2\bfx'\over d\xi'^2}\biggr)
   {d\tau'\over d\xi'}\biggl(1-\Bigl(\hat\bfk\cdot{d\bfx'\over d\eta'}\Bigr)^2\biggr)
\nonumber\\
&&~~~~~~+2\biggl(\Bigl({{\hat \bfk}\cdot{d^2\bfx\over d\xi^2}}\Bigr)
          \Bigl({{\hat \bfk}\cdot{d^3\bfx'\over d\xi'^3}}\Bigr)
          -{d^2\bfx\over d\xi^2}\cdot{d^3\bfx'\over d\xi'^3}\biggr)
   {d\tau\over d\xi}\biggl(1-\Bigl(\hat\bfk\cdot{d\bfx\over d\eta}\Bigr)^2\biggr)
\nonumber\\
&&~~~~~~-\biggl(\Bigl({{\hat \bfk}\cdot{d^3\bfx\over d\xi^3}}\Bigr)
          \Bigl({{\hat \bfk}\cdot{d\bfx'\over d\xi'}}\Bigr)
          -{d^3\bfx\over d\xi^3}\cdot{d\bfx'\over d\xi'}\biggr)
{d \over d\xi'}\biggl(
{d\tau'\over d\xi'}\biggl(1-\Bigl(\hat\bfk\cdot{d\bfx'\over d\eta'}\Bigr)^2\biggr)\biggr)
\nonumber\\
&&~~~~~~+\biggl(\Bigl({{\hat \bfk}\cdot{d\bfx\over d\xi}}\Bigr)
          \Bigl({{\hat \bfk}\cdot{d^3\bfx'\over d\xi'^3}}\Bigr)
          -{d\bfx\over d\xi}\cdot{d^3\bfx'\over d\xi'^3}\biggr)
 {d \over d\xi}\biggl(
  {d\tau\over d\xi}\biggl(1-\Bigl(\hat\bfk\cdot{d\bfx\over d\eta}\Bigr)^2\biggr)\biggr).
\end{eqnarray}
Then, we find
\begin{eqnarray}
E^{(1)}&=&
{e^2\hbar \over  (4\pi)^3\epsilon_0}\int d\Omega_{\hat\bfk} 
\int d\xi \int d\xi' {1\over \xi-\xi'}
\nonumber
\\
&&\times\Biggl\{
\biggl(\Bigl({{\hat \bfk}\cdot{d^3\bfx\over d\xi^3}}\Bigr)
          \Bigl({{\hat \bfk}\cdot{d^2\bfx'\over d\xi'^2}}\Bigr)
          -{d^3\bfx\over d\xi^3}\cdot{d^2\bfx'\over d\xi'^2}
         -\Bigl({{\hat \bfk}\cdot{d^2\bfx\over d\xi^2}}\Bigr)
          \Bigl({{\hat \bfk}\cdot{d^3\bfx'\over d\xi'^3}}\Bigr)
          +{d^2\bfx\over d\xi^2}\cdot{d^3\bfx'\over d\xi'^3}\biggr)
\nonumber\\&&~~~~~~\times
   \biggl(\hat\bfk\cdot{d\bfx\over d\eta} {d\tau\over d\eta}+
          \hat\bfk\cdot{d\bfx'\over d\eta'} {d\tau'\over d\eta'}\biggr)
\nonumber\\
&&~~~~+2\biggl(\Bigl({{\hat \bfk}\cdot{d^2\bfx\over d\xi^2}}\Bigr)
          \Bigl({{\hat \bfk}\cdot{d^2\bfx'\over d\xi'^2}}\Bigr)
          -{d^2\bfx\over d\xi^2}\cdot{d^2\bfx'\over d\xi'^2}\biggr)
   \biggl(
   {d\over d\xi}\Bigl(\hat\bfk\cdot{d\bfx\over d\eta} {d\tau\over d\eta}\Bigr)
-  {d\over d\xi'}\Bigl(\hat\bfk\cdot{d\bfx'\over d\eta'} {d\tau'\over d\eta'}\Bigr)
  \biggr)
\nonumber\\
&&~~~~+\biggl(\Bigl({{\hat \bfk}\cdot{d\bfx\over d\xi}}\Bigr)
          \Bigl({{\hat \bfk}\cdot{d^2\bfx'\over d\xi'^2}}\Bigr)
          -{d\bfx\over d\xi}\cdot{d^2\bfx'\over d\xi'^2}\biggr)
{d^2\over d\xi^2}\Bigl(\hat\bfk\cdot{d\bfx\over d\eta} {d\tau\over d\eta}\Bigr)
\nonumber\\
&&~~~~-\biggl(\Bigl({{\hat \bfk}\cdot{d^2\bfx\over d\xi^2}}\Bigr)
          \Bigl({{\hat \bfk}\cdot{d\bfx'\over d\xi'}}\Bigr)
          -{d^2\bfx\over d\xi^2}\cdot{d\bfx'\over d\xi'}\biggr)
{d^2\over d\xi'^2}\Bigl(\hat\bfk\cdot{d\bfx'\over d\eta'} 
{d\tau'\over d\eta'}\Bigr)
\nonumber\\
&&~~~~+\biggl(\Bigl({{\hat \bfk}\cdot{d^3\bfx\over d\xi^3}}\Bigr)
          \Bigl({{\hat \bfk}\cdot{d\bfx'\over d\xi'}}\Bigr)
          -{d^3\bfx\over d\xi^3}\cdot{d\bfx'\over d\xi'}\biggr)
{d\over d\xi'}\Bigl(\hat\bfk\cdot{d\bfx'\over d\eta'} {d\tau'\over d\eta'}\Bigr)
\nonumber\\
&&~~~~-\biggl(\Bigl({{\hat \bfk}\cdot{d\bfx\over d\xi}}\Bigr)
          \Bigl({{\hat \bfk}\cdot{d^3\bfx'\over d\xi'^3}}\Bigr)
          -{d\bfx\over d\xi}\cdot{d^3\bfx'\over d\xi'^3}\biggr)
{d\over d\xi}\Bigl(\hat\bfk\cdot{d\bfx\over d\eta} {d\tau\over d\eta}\Bigr)
\nonumber\\
&&~~~~+
2\biggl(\Bigl({{\hat \bfk}\cdot{d^3\bfx\over d\xi^3}}\Bigr)
          \Bigl({{\hat \bfk}\cdot{d^3\bfx'\over d\xi'^3}}\Bigr)
          -{d^3\bfx\over d\xi^3}\cdot{d^3\bfx'\over d\xi'^3}\biggr)
   \int_{\xi'}^\xi d\xi'' {d\tau''\over d\xi''}
   \biggl(1-\Bigl(\hat\bfk\cdot{d\bfx''\over d\eta''}\Bigr)^2\biggr)
\nonumber\\
&&~~~~-4\biggl(\Bigl({{\hat \bfk}\cdot{d^3\bfx\over d\xi^3}}\Bigr)
          \Bigl({{\hat \bfk}\cdot{d^2\bfx'\over d\xi'^2}}\Bigr)
          -{d^3\bfx\over d\xi^3}\cdot{d^2\bfx'\over d\xi'^2}\biggr)
   {d\tau'\over d\xi'}\biggl(1-\Bigl(\hat\bfk\cdot{d\bfx'\over d\eta'}\Bigr)^2\biggr)
\nonumber\\
&&~~~~+4\biggl(\Bigl({{\hat \bfk}\cdot{d^2\bfx\over d\xi^2}}\Bigr)
          \Bigl({{\hat \bfk}\cdot{d^3\bfx'\over d\xi'^3}}\Bigr)
          -{d^2\bfx\over d\xi^2}\cdot{d^3\bfx'\over d\xi'^3}\biggr)
   {d\tau\over d\xi}\biggl(1-\Bigl(\hat\bfk\cdot{d\bfx\over d\eta}\Bigr)^2\biggr)
\nonumber\\
&&~~~~-2\biggl(\Bigl({{\hat \bfk}\cdot{d^3\bfx\over d\xi^3}}\Bigr)
          \Bigl({{\hat \bfk}\cdot{d\bfx'\over d\xi'}}\Bigr)
          -{d^3\bfx\over d\xi^3}\cdot{d\bfx'\over d\xi'}\biggr)
{d \over d\xi'}\biggl(
{d\tau'\over d\xi'}\biggl(1-\Bigl(\hat\bfk\cdot{d\bfx'\over d\eta'}\Bigr)^2\biggr)\biggr)
\nonumber\\
&&~~~~+2\biggl(\Bigl({{\hat \bfk}\cdot{d\bfx\over d\xi}}\Bigr)
          \Bigl({{\hat \bfk}\cdot{d^3\bfx'\over d\xi'^3}}\Bigr)
          -{d\bfx\over d\xi}\cdot{d^3\bfx'\over d\xi'^3}\biggr)
 {d \over d\xi}\biggl(
  {d\tau\over d\xi}\biggl(1-\Bigl(\hat\bfk\cdot{d\bfx\over d\eta}
\Bigr)^2\biggr)\biggr)
\Biggl\}.
\label{Eeee1}
\end{eqnarray}
\def\vslash{{\backslash\hspace{-1.5mm}v}}
\def\bfv{{\bf v}}
From the definition of $\xi$ by Eq.~(\ref{dexfi}), we have
\begin{eqnarray}
{d\xi\over d\eta}=1-{\hat \bfk}\cdot{d\bfx\over d\eta}=1-\vslash,
\label{Eeee7}
\end{eqnarray}
where we defined $\vslash={\hat \bfk}\cdot \bfv$. Then, we also have
\begin{eqnarray}
&&{d\bfx\over d\xi}={\bfv\over 1-\vslash},
\\
&&{d^2\bfx\over d\xi^2}={\dot\bfv\over (1-\vslash)^2}+{\bfv\dot\vslash\over (1-\vslash)^3},
\\
&&{d^3\bfx\over d\xi^3}={\ddot\bfv\over (1-\vslash)^3}+{\bfv\ddot\vslash\over (1-\vslash)^4}
+{3\dot\bfv\dot\vslash\over (1-\vslash)^4}+{3\bfv\dot\vslash^2\over (1-\vslash)^5},
\end{eqnarray}
and
\begin{eqnarray}
&&{d\tau\over d\eta}={1\over \sqrt{(\bfp_i-e\bfA)^2+m^2}}
\label{dtaudeta},
\\
&&{d\over d\xi}\left({\vslash{d\tau\over d\eta}}\right)={1\over 1-\vslash}\left(
{\dot\vslash\over\sqrt{(\bfp_i-e\bfA)^2+m^2}}-{\vslash(\bfp_i-e\bfA)\cdot(-e\dot\bfA)\over\sqrt{(\bfp_i-e\bfA)^2+m^2}^3}
\right),
\\
&&{d^2\over d\xi^2}\left({\vslash{d\tau\over d\eta}}\right)={\dot\vslash\over (1-\vslash)^3}\left(
{\dot\vslash\over\sqrt{(\bfp_i-e\bfA)^2+m^2}}-{\vslash(\bfp_i-e\bfA)\cdot(-e\dot\bfA)\over\sqrt{(\bfp_i-e\bfA)^2+m^2}^3}
\right)
\nonumber\\
&&~~~~~~~~~~~+{1\over (1-\vslash)^2}\biggl(
{\ddot\vslash\over\sqrt{(\bfp_i-e\bfA)^2+m^2}}
+{3\vslash((\bfp_i-e\bfA)\cdot(-e\dot\bfA))^2\over\sqrt{(\bfp_i-e\bfA)^2+m^2}^5}
\nonumber
\\
&&~~~~~~~~~~~
-{2\dot\vslash(\bfp_i-e\bfA)\cdot(-e\dot\bfA)+\vslash(-e\dot\bfA)\cdot(-e\dot\bfA)+\vslash(\bfp_i-e\bfA)\cdot(-e\ddot\bfA)\over\sqrt{(\bfp_i-e\bfA)^2+m^2}^3}
\biggr).
\end{eqnarray}

In the limit of a non-relativistic motion of a charged particle, 
we use the following approximate formulas, 
\begin{eqnarray}
&&{d\bfx\over d\xi}\simeq{\bf v},
~~~~~~{d^2\bfx\over d\xi^2}\simeq\dot {\bf v},
~~~~~~{d^3\bfx\over d\xi^3}\simeq\ddot {\bf v},
\nonumber
\end{eqnarray}
\begin{eqnarray}
&&{d\tau\over d\eta}\simeq{1\over m},
~~~~~~{d\over d\xi}\left({\vslash{d\tau\over d\eta}}\right)\simeq
{\dot\vslash\over m},
~~~~~~{d^2\over d\xi^2}\left({\vslash{d\tau\over d\eta}}\right)\simeq
{\ddot\vslash^2\over m},
\nonumber
\end{eqnarray}
then, we have the following expression by neglecting sub-leading terms, 
\begin{eqnarray}
E^{(1)}&=&
{e^2\hbar \over  (4\pi)^3\epsilon_0}\int d\Omega_{\hat\bfk} 
\int dt \int dt' {1\over t-t'}
\nonumber
\\
&&\times{2\over m}\Biggl\{
(\dot\vslash\dot\vslash'-\dot\bfv\cdot\dot\bfv')(\dot\vslash-\dot\vslash')
+(\ddot\vslash\ddot\vslash'-\ddot\bfv\cdot\ddot\bfv')(t-t')
-2(\ddot\vslash\dot\vslash'-\ddot\bfv\cdot\dot\bfv')
+2(\dot\vslash\ddot\vslash'-\dot\bfv\cdot\ddot\bfv')\Biggl\}.
\label{Eeee1nonrela}
\end{eqnarray}
We also neglect the first term of the right-hand side of (\ref{Eeee1nonrela}),
which is of order of $\dot v^3$.
After the integration with respect to $\hat \bfk$, we have
\begin{eqnarray}
E^{(1)}&=&
{e^2\hbar \over  (4\pi)^2\epsilon_0}
\int dt \int dt' {4\over 3m}\Biggl\{
-\ddot\bfv\cdot\ddot\bfv'
+{2(\ddot\bfv\cdot\dot\bfv'-\dot\bfv\cdot\ddot\bfv')\over  t-t'}
\Biggl\}.
\label{Eeee1nonrelb}
\end{eqnarray}
The first term of the right-hand side of (\ref{Eeee1nonrelb})
gives no contribution by assuming that the acceleration is
zero at the boundaries of the time.  Then, Eq.~(\ref{EE1}) is
obtained. 

Now let us consider the case when the particle is moving 
with a relativistic speed in the
direction parallel to the electric field. 
We choose the $z$ axis parallel to this direction. 
Then, we may write $\bfA=(0,0,A(t))$, $\bfv=(0,0,v)$, and 
$\bfp_{\rm i}=(0,0,p_{\rm i})$. In this case, we have
\begin{eqnarray}
\dot v\simeq -{m^2e \dot A \over p_i^3},~~~~~~~
\ddot v\simeq -{m^2e \ddot A\over p_i^3}.
\nonumber
\end{eqnarray}
We also have
\begin{eqnarray}
&&{d^2z\over d\xi^2}\simeq
{-1\over (1-v\cos\theta)^3}{m^2e \dot A \over p_i^3},
~~~~~~
{d^3z\over d\xi^3}\simeq{-1\over (1-v\cos\theta)^4}
{m^2e \ddot A \over p_i^3},
\nonumber
\end{eqnarray}
\begin{eqnarray}
&&{d\tau\over d\eta}\simeq{1\over p_i},
~~~~~~{d\over d\xi}\left({\vslash{d\tau\over d\eta}}\right)\simeq
{v\cos\theta\over 1-v\cos\theta}{e\dot A\over p_i^2},
\nonumber
\end{eqnarray}
and
\begin{eqnarray}
&&{d^2\over d\xi^2}\left({\vslash{d\tau\over d\eta}}\right)\simeq
{v\cos\theta\over (1-v\cos\theta)^2}{e\ddot A\over p_i^2}
-{v\cos\theta\over (1-v\cos\theta)^3}{(e\dot A)^2m^2\over p_i^5},
\nonumber
\end{eqnarray}
in the limit of the relativistic motion. 
These approximate formulas yield the expression Eq.~(\ref{E1D})
at the leading order. 
In this derivation, we here note that the following term 
of the integration included in Eq.~(\ref{Eeee1}), 
\begin{eqnarray}
\int_{\xi'}^\xi d\xi'' {d\tau''\over d\xi''}
   \biggl(1-\Bigl(\hat\bfk\cdot{d\bfx''\over d\eta''}\Bigr)^2\biggr)
   &=&\int_{\xi'}^\xi d\xi''{d\tau''\over dt''}{dt''\over d\xi''}
\biggl(1-\vslash^2(t'')\biggr), 
\label{intterm}
\end{eqnarray}
gives no contribution at the leading order. 
Using Eqs.~(\ref{Eeee7}) and (\ref{dtaudeta}), the right-hand side of
Eq.~(\ref{intterm}) is written as
\begin{eqnarray}
\int_{\xi'}^\xi d\xi''{d\tau''\over dt''}{dt''\over d\xi''}
\biggl(1-\vslash^2(t'')\biggr)
   &=&\int_{\xi'}^\xi d\xi''{(1+\vslash(t''))\over \sqrt{(\bfp_i-e\bfA)^2+m^2}}.
\label{inttermb}
\end{eqnarray}
In the limit of the relativistic motion, the leading term is 
\begin{eqnarray}
\int_{\xi'}^\xi d\xi''{(1+\vslash(t''))\over \sqrt{(\bfp_i-e\bfA)^2+m^2}}
   &\simeq&{1\over p_i}
   \int_{\xi'}^\xi d\xi''\left(1+{{e\bf A}\cdot {\bf p_i}\over |p_i|^2}
\right)(1+\vslash(t''))
\nonumber
\\
   &\simeq&{1+\cos\theta\over p_i}\left(\xi-\xi'\right).
\label{inttermc}
\end{eqnarray}
The contribution of this term to $E^{(1)}$ is zero by 
assuming that the acceleration is zero at the boundaries 
of the time.

\section{Validity of WKB Approximation}
We consider the validity of using the WKB approximation, 
which breaks down when the background field varies rapidly.
The following condition is necessary to use 
the WKB approximation (e.g., \cite{BD}),
\begin{eqnarray}
 \frac{1}{2\Omega_{\bf p}^2}
\left|\frac{\ddot{\Omega}_{\bf p}}{\Omega_{\bf p}}-
\frac{3}{2}\frac{\dot{\Omega}_{\bf p}^2}{\Omega_{\bf p}^2}\right|
\ll1.
\end{eqnarray}
Using the expression (\ref{omegabfp}), this condition yields
\begin{eqnarray}
{\hbar^2\over 2((\bfp_i-e\bfA)^2+m^2)^3}\biggl|
{5\over 2}(e\dot\bfA\cdot(\bfp_i-e\bfA))^2
+((\bfp_i-e\bfA)^2+m^2)(e\ddot\bfA\cdot(\bfp_i-e\bfA)-(e\dot\bfA)^2)
\biggr|\ll1.
\label{WKBB}
\end{eqnarray}
In the relativistic limit, $|\bfp_i|\gg|e\bfA|,~m$, 
as is considered in section 3,
(\ref{WKBB}) reduces to
\begin{eqnarray}
{\hbar^2\over 2(\bfp_i^2)^3}\biggl|
{5\over 2}(e\dot\bfA\cdot \bfp_i)^2+\bfp_i^2(e\ddot\bfA\cdot\bfp_i)
\biggr|\ll1.
\label{WKBC}
\end{eqnarray}
In the case of the periodic electric field, $\dot A=-E_0\sin\omega t$, 
(\ref{WKBC}) requires
\begin{eqnarray}
{\hbar^2e^2E_0^2\over p_i^4}\ll1,
~~~{\rm and}~~~{\hbar^2 eE_0\omega \over p_i^3}\ll1,
\end{eqnarray}
which can be rewritten as
\begin{eqnarray}
\left(\hbar\omega\over p_i\right)^2\left(eE_0\over p_i\omega\right)^2\ll1,
~~~{\rm and}~~~
\left(\hbar\omega\over p_i\right)^2\left(eE_0\over p_i\omega\right)\ll1.
\label{WKBF}
\end{eqnarray}
We impose $eE_0/\omega\ll p_i$, as a condition of the limit of 
the relativistic motion. 
Then, the first inequality of (\ref{WKBF}) is satisfied when the 
second inequality of (\ref{WKBF}) is satisfied. 
Then, the condition 
required for the WKB approximation is written as,
\begin{eqnarray}
 1.3\times10^{-10}\biggl(\frac{\omega}{10^{15}{\rm s}^{-1}}\biggr)
\biggl(\frac{eE_0}{1\times 10^{15}{\rm eV/m}}\biggr)
  \biggl(\frac{p_{\rm i}c}{{\rm MeV}}\biggr)^{-3}
\ll1.
\end{eqnarray}
\end{appendix}

\end{document}